\documentclass[11pt]{article}

\usepackage{geometry}
\usepackage{booktabs}
\usepackage{float}
\usepackage[parfill]{parskip}
\usepackage{natbib}
\usepackage[nottoc,numbib]{tocbibind}
\usepackage[inline]{enumitem}

\usepackage{url}
\usepackage{hyperref}

\usepackage{authblk}

\usepackage[format=hang,labelfont=bf]{caption}
\usepackage[raggedright]{titlesec}

\usepackage{lscape}
\usepackage{multirow}

\usepackage[table]{xcolor}
\definecolor{lightgray}{gray}{0.95}

\title{A Guide to Comparing the Performance of VA Algorithms}

\author[1,2,*,$\dagger$]{Samuel J. Clark}

\affil[1]{Department of Sociology, The Ohio State University}
\affil[2]{MRC/Wits Rural Public Health and Health Transitions Research Unit (Agincourt), School of Public Health, Faculty of Health Sciences, University of the Witwatersrand}
\affil[*] {Contact: work@samclark.net}
\affil[$\dagger$] {Zehang (Richard) Li, Peter Byass, Martin Bratschi, and Philip Setel provided helpful discussion.}

\date{February 20, 2018}                                           

\begin{document}
\maketitle

\begin{abstract}
\noindent The literature comparing the performance of algorithms for assigning cause of death using verbal autopsy data is fractious and does not reach a consensus on which algorithms perform best, or even how to do the comparison.  This manuscript explains the challenges and suggests a way forward.  A universal challenge is the lack of standard training and testing data.  This limits meaningful comparisons between algorithms, and further, limits the ability of any algorithm to classify verbal autopsy deaths by cause in a way that is widely generalizable across regions and through time.  Verbal autopsy algorithms utilize a variety of information to describe the relationship between verbal autopsy symptoms and causes of death - called symptom-cause information (SCI).  A crowd sourced, public archive of SCI managed by the World Health Organization (WHO) is suggested as a way to address the lack of SCI for developing, testing, and comparing verbal autopsy coding algorithms, and additionally, as a way to ensure that algorithm-assigned causes of death are as accurate and comparable across regions and through time as possible.
\end{abstract}

\tableofcontents

\pagenumbering{arabic}

\section{Essential Components of VA Algorithms}

VA cause-assignment algorithms rely on three components:
\begin{enumerate*}[label=(\roman*)]
\item \textbf{VA data},
\item \textbf{symptom-cause information (SCI)}, and
\item \textbf{a logical algorithm that combines the two to identify cause-specific mortality fractions (CSMF)s and/or assign a likely cause of death to each death}
\end{enumerate*}.  
The SCI describes how VA symptoms are related to each cause -- either elicited directly physicians or in the form of a dataset of VAs and causes assigned using a different mechanism, e.g. medical records review (more on this below).  Additionally, algorithms produce different outputs. All create CSMFs, but some do not assign causes to individual deaths.  Among those that do, some assign one cause to each death while others assign a propensity or a probability to every cause for each death.  Those that assign a number to each cause report a summary of those numbers for each death, often the three causes with the largest number. 

The three components listed above -- data, SCI, and logic -- each affect the performance of VA algorithms.  It is possible to compare the effects of each by holding the other two constant.  Of obvious interest are the effects of the \textit{algorithm logic} and the \textit{SCI}.  In a comparison of this type the VA data must be held constant, and the algorithm logic and SCI varied independently -- i.e. one varied while holding the other constant.  Differences observed in causes assigned to individual deaths and in CSMFs can then be attributed to either the algorithm logic or the SCI.  Most published comparisons conflate the effects of logic and SCI by varying both together in unsystematic ways.  When that happens, differences in the causes assigned and CSMFs are due to differences in both the logic and SCI, and it is not possible to determine how much each component affects the differences.  \textit{In order to isolate the effect of algorithm logic, both the VA data and the SCI must be held constant and only the algorithm logic allowed to vary.}

Finally, the software that implements an algorithm often contains additional logic to either \textit{preprocess} the data (e.g. so that impossible symptom combinations are not included) or to \textit{postprocess} the results (e.g. to ensure that `impossible' causes are not reported, such as pregnancy-related deaths for males).  These software features are typically not reported in the publication that describes the method, but they can have important consequences for the real-world application of the algorithm.  Because of this, and the fact that transparency encourages both trust and innovation, it is important to have access to the source code for the software that implements an algorithm.

\section{Factors to Consider when Comparing the Performance of VA Algorithms}

Beyond the design of the comparison, additional factors affect the how the comparisons are done and how to interpret them.   

Algorithmic comparisons require VA data and SCI, and both are related to real deaths -- VA data are obtained from real deaths and SCI are often derived from real deaths that have causes assigned through a mechanism other than VA.  Because VA is most useful and most often applied to community deaths that occur outside of health facilities, \textit{it is important to know what types of deaths are included in the VA data and SCI}, either `community deaths' or `hospital deaths'.  Clearly, hospital deaths are less relevant to VA applications that target deaths occurring in the community. 

The algorithm logic is of obvious importance, as is the SCI (see above and below).  Beyond those, the structure of the test is important: does it use real data or is it a simulation study, what are the details of the metrics used to characterize and compare the performance of the algorithms, and how are causes and symptoms included in the test -- at their `natural' frequencies or resampled to over/under-represent certain symptoms or causes.  

Perhaps most important is whether the test deaths are part of the training data or completely different.  A common test design involves obtaining a large collection of deaths with both VA data and causes assigned through another mechanism.  All of the testing and comparisons are done using these `gold standard' deaths.  A sample of these deaths is used to generate the SCI for the algorithms, and then the algorithms are run on the remaining deaths, and the results compared.  This setup holds both the VA data and SCI constant and conducts tests on deaths that are on average indistinguishable from those used to generate the SCI.  Any differences observed in this setup are created by the algorithms and nothing else.  This characterizes how well each algorithm performs if an algorithm is applied to VA data that were not used to build the logic of the algorithm, and the results can be used to fairly compare algorithms without worrying about the effects of SCI.  However, this test does not indicate how well the algorithms perform when using different SCI or when applied to different deaths.  This is critical because most tests done using a framework like this are conducted on gold standard \textit{hospital} deaths.  In that case, the results say little about how the algorithms perform on community deaths.  At this time, there are no good gold standard datasets for community deaths, so it is not possible to do the test we really want.

Some published comparisons do compare the performance of algorithms on community deaths or the performance of algorithms to physician coding using community deaths.  In these cases there is no additional reference cause of death to which either algorithms or physician codes can be compared, so the results are  difficult to interpret.  If the physician codes are viewed as a reference, then the performance of algorithms can be assessed relative to physicians.

\section{Symptom-Cause Information (SCI)}

SCI describes how VA symptoms are related to each cause. SCI are different from `training data' because they can include both labeled data (traditional training data) and other forms of knowledge that inform the relationship between symptoms and causes.  There are effectively two commonly-used collections of SCI: 
\begin{enumerate*}[label=(\roman*)]
\item \textbf{the InterVA SCI} \citep{byass2012strengthening}, and
\item \textbf{the Population Health Metrics Research Consortium (PHMRC) gold standard dataset} \citep{murray2011population}
\end{enumerate*}.

The InterVA SCI was developed in an iterative way in several population surveillance research sites in South Africa, Ethiopia and Vietnam.  Consequently the InterVA SCI broadly relate to those specific populations and cannot be readily updated or adapted.  The PHMRC SCI are derived from the PHMRC gold standard dataset of deaths with VA and medically certified causes.  

The PHMRC dataset contains deaths that occurred in hospitals in six populations (Andhra Pradesh, India; Bohol, Philippines; Dar es Salaam, Tanzania; Mexico City, Mexico; Pemba Island, Tanzania; and Uttar Pradesh, India) in the years leading up to 2011.  Consequently the PHMRC SCI describes the relationships between VA symptoms and causes of death in hospitals in those six specific populations during the late 2000s. These relationships have likely shifted with time and are likely of limited applicability in other geographic locations.  

SCI relate VA symptoms to causes of death, and consequently, a specific SCI comes with a defined symptom list and a defined cause list.  The InterVA SCI symptom and cause lists correspond to the WHO 2012 and WHO 2016 standard VA tools \citep{whoStandardTools2012,whoStandardTools2016}, while the PHMRC SCI symptom and cause lists correspond to those defined by the original PHMRC study \citep{murray2011population}.  Some algorithms use subsets of the symptom and causes lists for a given SCI.

\section{VA Algorithms}

There are six computational algorithms that have been seriously proposed and/or used widely.  Physicians can also be thought of as an algorithm, and there are additional algorithms that have been proposed but not considered seriously for real-world use, e.g. several earlier propositions by the Institute for Health Metrics and Evaluation (IHME), e.g. `Random Forest' and `Simplified Symptom Pattern'.  

The \textbf{InterVA} algorithm \citep[e.g.][]{byass2012strengthening,Fottrell2007,ISI:000208221400021,ISI:000235757400013,ISI:000297197000031} was developed and refined by Peter Byass and his colleagues over many years and is currently in version 5, to be released soon.  InterVA uses a reduced version of Bayes rule and SCI in the form of conditional probabilities of experiencing a symptom given a specific cause of death: $\Pr(s|c)$.  The conditional probabilities are elicited from physicians, and in the latest version of InterVA (InterVA-5), the PHMRC gold standard dataset also informs some of the conditional probabilities.  InterVA identifies a propensity for each cause of death for each death and derives CSMFs by adding up those propensities across all deaths in a VA dataset.  There is open-source software available for InterVA4.
\begin{itemize}
\item Open source windows software: \url{http://interva.net}
\item Open-source \texttt{R} package: \url{https://cran.r-project.org/package=InterVA4}
\end{itemize}

The \textbf{Tariff} algorithm \citep{james2011performance} was developed by IHME and has been refined as the SmartVA algorithm \citep{serina2015improving,serina2015shortened} that requires fewer symptoms than Tariff and can classify some causes better than Tariff.  The Tariff family of algorithms use an additive tariff score to rank causes for each death.  SCI for the Tariff algorithms are individual tariffs that correspond to the strength of association between particular symptoms and causes in the PHMRC gold standard dataset.  The SmartVA-Analyze computer software adds some `expert opinion' to the SCI derived from the PHMRC gold standard dataset.  The Tariff methods identify the single cause with the highest rank for each death in a VA dataset and add those up to derive CSMFs.  There is free proprietary software for SmartVA-Analyze available from IHME and an open-source \texttt{R} package to implement Tariff.
\begin{itemize}
\item SmartVA-Analyze: \url{http://www.healthdata.org/verbal-autopsy/tools}
\item Open-source Tariff \texttt{R} package: \url{https://cran.r-project.org/package=Tariff}
\end{itemize}

The \textbf{InSilicoVA} algorithm \citep{mccormick2016probabilistic,mccormick2016probabilisticSupplement} was developed recently by our group of demographers and statisticians at the University of Washington and The Ohio State University.  InSilicoVA is a statistical algorithm based on Bayes rule that jointly identifies the most likely cause assignments for each death and the corresponding CSMFs for all deaths in a VA dataset.   InSilicoVA can use a variety of SCI including both the InterVA and PHMRC SCI and others if they are available.  InSilicoVA assesses uncertainty (e.g. provides credible intervals) in both individual cause assignments and CSMFs and is the only algorithm to do this.  There is an \texttt{R} package for InSilicoVA and reproducibility code available from the authors.
\begin{itemize}
\item Open-source InSilicoVA \texttt{R} package: \url{https://cran.r-project.org/package=InSilicoVA}
\end{itemize}

The \textbf{NBC} (naive Bayes classifier) algorithm \citep{miasnikof2015naive} was developed by a group of statisticians and epidemiologists at the University of Toronto.  NBC is a direct application of Bayes rule as a standard naive Bayes classifier.  As presented by its developers, NBC uses custom SCI derived by the Toronto group related to deaths in India.  There is an \texttt{R} package available that accepts training data and VA data so that the method can use a variety of SCI.
\begin{itemize}
\item Open-source NBC \texttt{R} package: \url{https://cran.r-project.org/package=nbc4va}
\end{itemize}

The \textbf{King-Lu} algorithm \citep{king2008verbal,king2010designing} was developed by King and Lu and uses Bayes rule to estimate only the the CSMFs using constrained optimization. The algorithm itself does not perform individual classification, but the authors describe a possible two-stage approach. The King-Lu method also relies on the SCI from training data. Uniquely, the King-Lu method considers the SCI described by groups of symptoms, instead of each symptom alone. Therefore it must use high quality gold standard deaths, preferably from the same population as the VA deaths, or otherwise researchers need to know which small groups of symptoms to use. This makes it impossible to use King-Lu in a situation without gold standard deaths.  There is an \texttt{R} package for the King-Lu method.
\begin{itemize}
\item Open-source King-Lu \texttt{R} package: \url{https://gking.harvard.edu/va}
\end{itemize}

There is an open-source \texttt{R} package called `openVA' that runs all of the open-source \texttt{R} packages listed above, except King-Lu, in a nice environment.
\begin{itemize}
\item Open-source openVA \texttt{R} package: \url{https://cran.r-project.org/package=openVA}
\end{itemize}

Table \ref{tab:algs} summarizes the properties of these six algorithms and physician coding.  In addition to what SCI they use, the table identifies what outputs are produced and whether or not the method is able to assess uncertainty in those outputs.  There is a column labeled `Independent Symptoms' that communicates whether or not the algorithm makes an assumption of independence among symptoms given cause.  This assumption makes the logic of all of the algorithms much simpler but effectively ignores information contained in the co-occurrence or co-absence of symptoms.  Finally, the table lists what type of software is available for each algorithm.

\begin{landscape}
\begin{table}
\captionsetup{format=plain,font=normalsize,margin=0cm,justification=justified}
\caption{\textbf{VA Algorithms.} \textit{SCI} is `symptom-cause information'; \textit{PHMRC} is `Population Health Metrics Research Consortium gold standard VA dataset' ; \textit{Outputs} are `individual COD', `CSMF', or `both'; \textit{O-S} is `open-source'; \textit{\texttt{R}} is `\texttt{R} statistical programming environment'.}
\begin{center}
\begin{tabular}{ l c c c c c c c c c c c }
\toprule
& \multirow{2}{*}[-5pt]{\begin{tabular}[x]{@{}c@{}}Fundamental\\Logic\end{tabular}} & & \multicolumn{3}{c}{Alternate SCI} & & & \multirow{2}{*}[-5pt]{\begin{tabular}[x]{@{}c@{}}Indep.\\Symptoms\end{tabular}} & \multicolumn{3}{c}{Software} \\
\cmidrule(lr){4-6} \cmidrule(lr){10-12} 
Algorithm &  & SCI & PHMRC & InterVA & Other & Outputs & Uncertainty & & Free & O-S & Platforms \\
\midrule
							InterVA 			& $\sim$Bayes rule 	& InterVA  		& No 	& Yes 	& No 	& Both 			& No 	& Yes 			& Yes 	& Yes 	& All (\texttt{R})	\\
\rowcolor{gray!25} Tariff 			& tariff scores 				& PHMRC			& Yes 	& No		& No 	& Both 		& No 	& Yes 			& Yes 	& No 	& Windows \\
							SmartVA 		& tariff scores 				& PHMRC			& Yes 	& No		& No 	& Both 		& No 	& Yes 			& Yes 	& No 	& Windows \\
\rowcolor{gray!25} InSilicoVA 	& Bayes rule+ 				& Various			& Yes 	& Yes 	& Yes 	& Both 		& Yes 	& Yes 			& Yes 	& Yes 	& All (\texttt{R}) \\
							NBC 			& Bayes rule 				& Various			& Yes 	& Yes 	& Yes 	& Both 		& No 	& Yes 			& Yes 	& Yes 	& All (\texttt{R}) \\
\rowcolor{gray!25} King-Lu 		& Bayes rule+				& Various			& Yes 	& No 	& Yes 	& CSMF	& No 	& $\sim$No 	& Yes 	& Yes 	& All (\texttt{R}) \\
							Physicians 	& Various						& \begin{tabular}[x]{@{}c@{}}School \& \\ Experience\end{tabular}			
																											& No 	& No 	& Yes 	& Both 		& No		& No 			& -NA- 	& -NA- 	& -NA-\\
\bottomrule
\end{tabular}
\end{center}
\label{tab:algs}
\end{table}%
\end{landscape}

\section{Recently Published Comparisons of the Performance of VA Algorithms}

There are three recent publications that present comprehensive comparisons of the performance of commonly used algorithms.  

\textbf{Performance of four computer-coded verbal autopsy methods for cause of death assignment compared with physician coding on 24,000 deaths in low- and middle-income countries} \citep{desai2014performance}.

\begin{quote}
\textbf{Background:} Physician-coded verbal autopsy (PCVA) is the most widely used method to determine causes of death (CODs) in countries where medical certification of death is uncommon. Computer-coded verbal autopsy (CCVA) methods have been proposed as a faster and cheaper alternative to PCVA, though they have not been widely compared to PCVA or to each other. 

\textbf{Methods:} We compared the performance of open-source random forest, open-source tariff method, InterVA-4, and the King-Lu method to PCVA on five datasets comprising over 24,000 verbal autopsies from low- and middle-income countries. Metrics to assess performance were positive predictive value and partial chance-corrected concordance at the individual level, and cause-specific mortality fraction accuracy and cause-specific mortality fraction error at the population level. 

\textbf{Results:} The positive predictive value for the most probable COD predicted by the four CCVA methods averaged about 43\% to 44\% across the datasets. The average positive predictive value improved for the top three most probable CODs, with greater improvements for open-source random forest (69\%) and open-source tariff method (68\%) than for InterVA-4 (62\%). The average partial chance-corrected concordance for the most probable COD predicted by the open-source random forest, open-source tariff method and InterVA-4 were 41\%, 40\% and 41\%, respectively, with better results for the top three most probable CODs. Performance generally improved with larger datasets. At the population level, the King-Lu method had the highest average cause-specific mortality fraction accuracy across all five datasets (91\%), followed by InterVA-4 (72\% across three datasets), open-source random forest (71\%) and open-source tariff method (54\%). 

\textbf{Conclusions:} On an individual level, no single method was able to replicate the physician assignment of COD more than about half the time. At the population level, the King-Lu method was the best method to estimate cause-specific mortality fractions, though it does not assign individual CODs. Future testing should focus on combining different computer-coded verbal autopsy tools, paired with PCVA strengths. This includes using open-source tools applied to larger and varied datasets (especially those including a random sample of deaths drawn from the population), so as to establish the performance for age- and sex-specific CODs.

\textit{Abstract verbatim from \cite{desai2014performance}}
\end{quote} 

\textbf{Using verbal autopsy to measure causes of death: the comparative performance of existing methods'} \citep{murray2014using}:

\begin{quote}
\textbf{Background:} Monitoring progress with disease and injury reduction in many populations will require widespread use of verbal autopsy (VA). Multiple methods have been developed for assigning cause of death from a VA but their application is restricted by uncertainty about their reliability. 

\textbf{Methods:} We investigated the validity of five automated VA methods for assigning cause of death: InterVA-4, Random Forest (RF), Simplified Symptom Pattern (SSP), Tariff method (Tariff), and King-Lu (KL), in addition to physician review of VA forms (PCVA), based on 12,535 cases from diverse populations for which the true cause of death had been reliably established. For adults, children, neonates and stillbirths, performance was assessed separately for individuals using sensitivity, specificity, Kappa, and chance-corrected concordance (CCC) and for populations using cause specific mortality fraction (CSMF) accuracy, with and without additional diagnostic information from prior contact with health services. A total of 500 train-test splits were used to ensure that results are robust to variation in the underlying cause of death distribution. 

\textbf{Results:} Three automated diagnostic methods, Tariff, SSP, and RF, but not InterVA-4, performed better than physician review in all age groups, study sites, and for the majority of causes of death studied. For adults, CSMF accuracy ranged from 0.764 to 0.770, compared with 0.680 for PCVA and 0.625 for InterVA; CCC varied from 49.2\% to 54.1\%, compared with 42.2\% for PCVA, and 23.8\% for InterVA. For children, CSMF accuracy was 0.783 for Tariff, 0.678 for PCVA, and 0.520 for InterVA; CCC was 52.5\% for Tariff, 44.5\% for PCVA, and 30.3\% for InterVA. For neonates, CSMF accuracy was 0.817 for Tariff, 0.719 for PCVA, and 0.629 for InterVA; CCC varied from 47.3\% to 50.3\% for the three automated methods, 29.3\% for PCVA, and 19.4\% for InterVA. The method with the highest sensitivity for a specific cause varied by cause.  

\textbf{Conclusions:} Physician review of verbal autopsy questionnaires is less accurate than automated methods in determining both individual and population causes of death. Overall, Tariff performs as well or better than other methods and should be widely applied in routine mortality surveillance systems with poor cause of death certification practices.

\textit{Abstract verbatim from \cite{murray2014using}}
\end{quote}

\textbf{Probabilistic Cause-of-Death Assignment Using Verbal Autopsies} \citep{mccormick2016probabilistic,mccormick2016probabilisticSupplement}:

\begin{quote}
In regions without complete-coverage civil registration and vital statistics systems there is uncertainty about even the most basic demographic indicators. In such regions, the majority of deaths occur outside hospitals and are not recorded. Worldwide, fewer than one-third of deaths are assigned a cause, with the least information available from the most impoverished nations. In populations like this, verbal autopsy (VA) is a commonly used tool to assess cause of death and estimate cause-specific mortality rates and the distribution of deaths by cause. VA uses an interview with caregivers of the decedent to elicit data describing the signs and symptoms leading up to the death. This article develops a new statistical tool known as InSilicoVA to classify cause of death using information acquired through VA. InSilicoVA shares uncertainty between cause of death assignments for specific individuals and the distribution of deaths by cause across the population. Using side-by-side comparisons with both observed and simulated data, we demonstrate that InSilicoVA has distinct advantages compared to currently available methods. Supplementary materials for this article are available online.

\textit{Abstract verbatim from \cite{mccormick2016probabilistic}}
\end{quote}

The InSilicoVA publication presents a number of results:
\begin{itemize}
\item Development of the InSilicoVA algorithm.
\item A simulation study of the performance of InSilicoVA.
\item A method to de-bias physician-coded VAs.
\item An application of InSilicoVA to community deaths with a comparison to InterVA.
\item An application of InSilicoVA that includes both standard SCI and physician-coded deaths that dramatically improves the performance of InSilicoVA.
\item An SCI-controlled comparison of the performance of algorithms including InSilicoVA, InterVA, Tariff, and King-Lu.
\end{itemize}
The overall conclusion is that InSilicoVA performs well and generally better than the other algorithms under a variety of tests (simulation, cross validation using gold standard deaths, and application to community deaths with and without additional physician-coded deaths as auxiliary SCI) and can be extended easily to use a variety of SCI and additional techniques to improve performance.  

In the SCI-controlled comparison, InSilicoVA consistently outperformed the other algorithms, but all algorithms performed much less well when tested on deaths from PHMRC sites that were not included to create their SCI.  \textit{This clearly demonstrates the need for much better SCI that is representative of community deaths in a wide variety of places and is updated through time, and the fact that the PHMRC gold standard dataset is not sufficient as a general purpose, generalizable source of SCI.}  

InSilicoVA is the only algorithm that quantifies uncertainty and the InSilicoVA publication is the only publication describing an algorithm for which full replication code exists (by request from the authors).

Table \ref{tab:comparisons} presents a very brief summary of the comparisons conducted and conclusions reached by each set of authors.

A number of other publications address the functioning and performance of VA algorithms and related issues: e.g.  \cite{
aleksandrowicz2014performance,
bauni2011validating,
byass2012strengthening,
byass2014usefulness,
flaxman2011direct,
flaxman2015measuring,
Fottrell2007,
fottrell2010verbal,
fottrell2011probabilistic,
garenne2014prospects,
ISI:000208221400021,
ISI:000235757400013,
ISI:000297197000031,
james2011performance,
jha2014reliable,
king2008verbal,
king2010designing,
leitao2014comparison,
lozano2011performance,
miasnikof2015naive,
mohapatra2016level,
murray2011population,
serina2015improving,
serina2015shortened,
whoStandardTools2012,
whoStandardTools2016}

\begin{landscape}
\begin{table}
\captionsetup{format=plain,font=normalsize,margin=0cm,justification=justified}
\caption{\textbf{Recent Published Comparisons of VA Algorithm Performance.} \textit{Def.} is `Default SCI'; \textit{Cntr.} is `Controlled (i.e. identical) SCI'.  \textit{Code} communicates whether or note replication code is available for the publication.  Only algorithms listed in the text are summarized in this table.}
\begin{center}
\begin{tabular}{ >{\raggedright}m{3cm} >{\raggedright}m{4cm} c c c c >{\raggedright\arraybackslash}m{8cm} }
\toprule
& & \multicolumn{2}{c}{SCI} &  &  \\ 
\cmidrule(lr){3-4} 
Reference & Compared & Def. & Cntr. & Death Type & Code & \multicolumn{1}{c}{Conclusion \& Comments} \\
\midrule
\cite{desai2014performance} & Physician codes compared to Tariff, InterVA4, and King-Lu  & Yes & No & Community & No  
& Algorithm-assigned causes compared to physician-assigned causes.  All algorithms using default SCI.  Algorithms performed poorly at replicating physician-assigned causes.  Tariff performed slightly better than InterVA4. 
\\
\rowcolor{gray!25} \cite{murray2014using}	& Algorithms to each other: InterVA4, Tariff, and King-Lu	& Yes & No 	& Hospital & No 
& Algorithm and physician-assigned causes compared.  All algorithms used default SCI and comparisons down within PHMRC gold standard data set.  Tariff performed better than physicians; InterVA4 performed worse than physicians.
\\
\cite{mccormick2016probabilistic,mccormick2016probabilisticSupplement} 	& Algorithms to each other: InSilicoVA, InterVA4, Tariff, King-Lu 	& Yes & Yes 	& Both 	& Yes 
& Algorithm-assigned causes compared.  All algorithms used SCI derived from the PHMRC gold standard dataset. All algorithms trained using data from all or single sites within the PHMRC data.  All algorithms performed much less well when trained and tested on deaths from different sites within the PHMRC data.  In general InSilicoVA performed better than all other algorithms, and InterVA4 performed better than Tariff.
\\
\bottomrule
\end{tabular}
\end{center}
\label{tab:comparisons}
\end{table}%
\end{landscape}

\section{Brief Conclusion}

\subsection{Performance of Algorithms}

The published evidence comparing the performance of VA algorithms does not present a consensus opinion that clearly identifies the `best' algorithm.  However, several clear conclusions can be drawn:
\begin{itemize}
\item With regard to underlying logic, the consensus is that some form of Bayes rule is the most fruitful approach, and with the sole exception of the Tariff family of methods, all algorithms use Bayes rule.
\item Except for \cite{mccormick2016probabilistic}, all published comparisons conflate the effects of algorithm logic and SCI, and that makes their results difficult to interpret.
\item The \cite{mccormick2016probabilistic} publication makes it clear that the effects of SCI can far outweigh differences in logic (see the SCI-controlled comparison done using PHMRC gold standard data in the Supplementary Materials for the publication).  Following directly from this, it is clear that \textit{a concerted effort must be made to significantly and rapidly improve the SCI available for VA algorithm developers and VA users.}
\item A standard set of VA deaths and SCI need to be developed to support future comparisons of the performance of VA methods.  This will ensure that variation in the test deaths and SCI do not affect the comparisons done by different VA algorithm research groups.
\item A standard comparison approach needs to be developed and utilized in all VA algorithm comparisons.
\item A standard set of VA algorithm performance metrics needs to be determined and reported in all comparisons of VA algorithms.
\end{itemize}

\subsection{SCI Archive -- an Approach to Improved SCI}

Because SCI relates VA symptoms to causes, the easiest and most reliable way to obtain SCI is from deaths with VA and a cause(s) assigned through an independent mechanism. These `labeled' deaths can be processed to produce a quantitative description of how each symptom or group of symptoms is related to each cause.  This typically consists of conditional probabilities of observing a symptom or group of symptoms with a specific cause $\Pr(s|c)$ or $\Pr({s_1, \dots, s_n}|c)$.  The independent mechanism for assigning causes to SCI deaths can be 1) a form of traditional medical certification, including medical record review and/or medical autopsy, 2) traditional physician-assigned causes using the VA data, or 3) a variety of mixtures of the two with perhaps additional information from clinical records and/or minimally-invasive tissue samples.

To allow automated cause assignment methods to be used with confidence in routine mortality surveillance in multiple settings and as time progresses, it is necessary to have SCI based on deaths from a wide variety of settings that accumulate continuously.  This variety can be achieved by continuously pooling deaths with VA and an independently-assigned cause(s) from various settings and updating SCI based on those deaths on a regular schedule.  Then by definition the resulting SCI are both widely representative and continuously updated.  

\textit{The practical implementation of such an idea is a centralized archive of deaths with VA and an independently-assigned cause(s) that continuously accepts deaths from a widely dispersed group of collaborators and periodically updates and disseminates SCI based on those deaths -- \textbf{the SCI Archive}.}   This is a straightforward idea, but actually implementing it will require working through some practical details.
\begin{itemize}
\item Ethical/legal concerns: privacy, data ownership, etc.
\item Data definitions for the VA symptoms and cause lists.
\item Quality assurance metrics for both the VAs and alternative cause assignment mechanisms.
\item Accessibility and dissemination formats, frequencies, etc.
\item Metadata definitions.
\item De-biasing data requirements for traditional physician-assigned causes (and perhaps others), see \citep{mccormick2016probabilistic}.
\item Computing/data management design and maintenance plans.
\item Physical computing infrastructure requirements.
\item Human resource needs, setup and maintenance phases.
\end{itemize}

A common desire among VA users in routine surveillance settings is to have automated cause-assignment algorithms `replicate what \textit{our} physicians would do' with VA data.  This replication can be achieved by using SCI based on deaths with physician-coded VAs representative of the area of VA implementation.  Physicians code VA deaths with their own idiosyncratic bias, and that is one of the primary motivations for developing algorithms.  This bias can be quantified and eliminated in the preparation of SCI from physician-coded VAs provided two conditions are met \citep{mccormick2016probabilistic}:
\begin{enumerate}
\item each VA death is read and coded by at least two physicians and 
\item each physician reads and codes many VAs.
\end{enumerate}
Physician-coded VAs that meet these conditions are `de-biased physician coded VAs' or DBPCVA.  

A potential way to quickly get the the SCI Archive up and running would be to start with DBPCVA deaths contributed by any VA user willing to provide them.  The resulting SCI would enable automated cause-assignment algorithms to assign causes in a way that replicates what non-biased physicians would do.  The WHO Working Group on VA is exploring the idea of setting up an SCI archive of this type at the WHO.

\newpage
\bibliographystyle{demography}
\bibliography{VA-Algorithm-Comparison}

\begin{thebibliography}{30}
\def\enquote#1{``#1''}
\expandafter\ifx\csname natexlab\endcsname\relax\def\natexlab#1{#1}\fi
\expandafter\ifx\csname url\endcsname\relax
  \def\url#1{{\tt #1}}\fi
\expandafter\ifx\csname urlprefix\endcsname\relax\def\urlprefix{URL }\fi

\bibitem[{Aleksandrowicz et~al.(2014)Aleksandrowicz, Malhotra, Dikshit, Gupta,
  Kumar, Sheth, Rathi, Suraweera, Miasnikof, Jotkar
  et~al.}]{aleksandrowicz2014performance}
Aleksandrowicz, L., V.~Malhotra, R.~Dikshit, P.~C. Gupta, R.~Kumar, J.~Sheth,
  S.~K. Rathi, W.~Suraweera, P.~Miasnikof, R.~Jotkar et~al. 2014.
  \enquote{Performance criteria for verbal autopsy-based systems to estimate
  national causes of death: development and application to the Indian Million
  Death Study.} {\em BMC medicine\/} 12(1):21.

\bibitem[{Bauni et~al.(2011)Bauni, Ndila, Mochamah, Nyutu, Matata, Ondieki,
  Mambo, Mutinda, Tsofa, Maitha et~al.}]{bauni2011validating}
Bauni, E., C.~Ndila, G.~Mochamah, G.~Nyutu, L.~Matata, C.~Ondieki, B.~Mambo,
  M.~Mutinda, B.~Tsofa, E.~Maitha et~al. 2011. \enquote{Validating
  physician-certified verbal autopsy and probabilistic modeling (InterVA)
  approaches to verbal autopsy interpretation using hospital causes of adult
  deaths.} {\em Population health metrics\/} 9(1):49.

\bibitem[{Byass(2014)}]{byass2014usefulness}
Byass, P. 2014. \enquote{Usefulness of the Population Health Metrics Research
  Consortium gold standard verbal autopsy data for general verbal autopsy
  methods.} {\em BMC medicine\/} 12(1):23.

\bibitem[{Byass et~al.(2012)Byass, Chandramohan, Clark, D'ambruoso, Fottrell,
  Graham, Herbst, Hodgson, Hounton, Kahn et~al.}]{byass2012strengthening}
Byass, P., D.~Chandramohan, S.~J. Clark, L.~D'ambruoso, E.~Fottrell, W.~J.
  Graham, A.~J. Herbst, A.~Hodgson, S.~Hounton, K.~Kahn et~al. 2012.
  \enquote{Strengthening standardised interpretation of verbal autopsy data:
  the new InterVA-4 tool.} {\em Global health action\/} 5(1):19281.

\bibitem[{Desai et~al.(2014)Desai, Aleksandrowicz, Miasnikof, Lu, Leitao,
  Byass, Tollman, Mee, Alam, Rathi et~al.}]{desai2014performance}
Desai, N., L.~Aleksandrowicz, P.~Miasnikof, Y.~Lu, J.~Leitao, P.~Byass,
  S.~Tollman, P.~Mee, D.~Alam, S.~K. Rathi et~al. 2014. \enquote{Performance of
  four computer-coded verbal autopsy methods for cause of death assignment
  compared with physician coding on 24,000 deaths in low-and middle-income
  countries.} {\em BMC medicine\/} 12(1):20.

\bibitem[{Fantahun et~al.({2006})Fantahun, Fottrell, Berhane, Wall, Hogberg and
  Byass}]{ISI:000235757400013}
Fantahun, M., E.~Fottrell, Y.~Berhane, S.~Wall, U.~Hogberg and P.~Byass.
  {2006}. \enquote{{Assessing a new approach to verbal autopsy interpretation
  in a rural Ethiopian community: the InterVA model}.} {\em {Bulletin of the
  World Health Organization}\/} {84}({3}):{204--210}.

\bibitem[{Flaxman et~al.(2015)Flaxman, Serina, Hernandez, Murray, Riley and
  Lopez}]{flaxman2015measuring}
Flaxman, A.~D., P.~T. Serina, B.~Hernandez, C.~J. Murray, I.~Riley and A.~D.
  Lopez. 2015. \enquote{Measuring causes of death in populations: a new metric
  that corrects cause-specific mortality fractions for chance.} {\em Population
  health metrics\/} 13(1):28.

\bibitem[{Flaxman et~al.(2011)Flaxman, Vahdatpour, James, Birnbaum and
  Murray}]{flaxman2011direct}
Flaxman, A.~D., A.~Vahdatpour, S.~L. James, J.~K. Birnbaum and C.~J. Murray.
  2011. \enquote{Direct estimation of cause-specific mortality fractions from
  verbal autopsies: multisite validation study using clinical diagnostic gold
  standards.} {\em Population health metrics\/} 9(1):35.

\bibitem[{Fottrell and Byass(2010)}]{fottrell2010verbal}
Fottrell, E. and P.~Byass. 2010. \enquote{Verbal autopsy: methods in
  transition.} {\em Epidemiologic reviews\/} 32(1):38--55.

\bibitem[{Fottrell et~al.(2007)Fottrell, Byass, Ouedraogo, Tamini, Gbangou,
  Sombie, Hogberg, Witten, Bhattacharya, Desta, Deganus, Tornui, Fitzmaurice,
  Meda and Graham}]{Fottrell2007}
Fottrell, E., P.~Byass, T.~Ouedraogo, C.~Tamini, A.~Gbangou, I.~Sombie,
  U.~Hogberg, K.~Witten, S.~Bhattacharya, T.~Desta, S.~Deganus, J.~Tornui,
  A.~Fitzmaurice, N.~Meda and W.~Graham. 2007. \enquote{Revealing the burden of
  maternal mortality: a probabilistic model for determining pregnancy-related
  causes of death from verbal autopsies.} {\em Population Health Metrics\/}
  5(1):1. \urlprefix\url{http://www.pophealthmetrics.com/content/5/1/1}.

\bibitem[{Fottrell et~al.({2011})Fottrell, Kahn, Tollman and
  Byass}]{ISI:000297197000031}
Fottrell, E., K.~Kahn, S.~Tollman and P.~Byass. {2011}. \enquote{{Probabilistic
  Methods for Verbal Autopsy Interpretation: InterVA Robustness in Relation to
  Variations in A Priori Probabilities}.} {\em {PLoS ONE}\/} {6}({11}).

\bibitem[{Fottrell et~al.(2011)Fottrell, Kahn, Tollman and
  Byass}]{fottrell2011probabilistic}
---{}---{}---. 2011. \enquote{Probabilistic methods for verbal autopsy
  interpretation: InterVA robustness in relation to variations in a priori
  probabilities.} {\em PLoS One\/} 6(11):e27200.

\bibitem[{Garenne(2014)}]{garenne2014prospects}
Garenne, M. 2014. \enquote{Prospects for automated diagnosis of verbal
  autopsies.} {\em BMC medicine\/} 12(1):18.

\bibitem[{James et~al.(2011)James, Flaxman and Murray}]{james2011performance}
James, S.~L., A.~D. Flaxman and C.~J. Murray. 2011. \enquote{Performance of the
  Tariff Method: validation of a simple additive algorithm for analysis of
  verbal autopsies.} {\em Population Health Metrics\/} 9(1):31.

\bibitem[{Jha(2014)}]{jha2014reliable}
Jha, P. 2014. \enquote{Reliable direct measurement of causes of death in
  low-and middle-income countries.} {\em BMC medicine\/} 12(1):19.

\bibitem[{King et~al.(2010)King, Lu and Shibuya}]{king2010designing}
King, G., Y.~Lu and K.~Shibuya. 2010. \enquote{Designing verbal autopsy
  studies.} {\em Population Health Metrics\/} 8(1):19.

\bibitem[{King et~al.(2008)King, Lu et~al.}]{king2008verbal}
King, G., Y.~Lu et~al. 2008. \enquote{Verbal autopsy methods with multiple
  causes of death.} {\em Statistical Science\/} 23(1):78--91.

\bibitem[{Leitao et~al.(2014)Leitao, Desai, Aleksandrowicz, Byass, Miasnikof,
  Tollman, Alam, Lu, Rathi, Singh et~al.}]{leitao2014comparison}
Leitao, J., N.~Desai, L.~Aleksandrowicz, P.~Byass, P.~Miasnikof, S.~Tollman,
  D.~Alam, Y.~Lu, S.~K. Rathi, A.~Singh et~al. 2014. \enquote{Comparison of
  physician-certified verbal autopsy with computer-coded verbal autopsy for
  cause of death assignment in hospitalized patients in low-and middle-income
  countries: systematic review.} {\em BMC medicine\/} 12(1):22.

\bibitem[{Lozano et~al.(2011)Lozano, Lopez, Atkinson, Naghavi, Flaxman and
  Murray}]{lozano2011performance}
Lozano, R., A.~D. Lopez, C.~Atkinson, M.~Naghavi, A.~D. Flaxman and C.~J.
  Murray. 2011. \enquote{Performance of physician-certified verbal autopsies:
  multisite validation study using clinical diagnostic gold standards.} {\em
  Population Health Metrics\/} 9(1):32.

\bibitem[{McCormick et~al.(2016{\natexlab{a}})McCormick, Li, Calvert, Crampin,
  Kahn and Clark}]{mccormick2016probabilistic}
McCormick, T.~H., Z.~R. Li, C.~Calvert, A.~C. Crampin, K.~Kahn and S.~J. Clark.
  2016{\natexlab{a}}. \enquote{Probabilistic cause-of-death assignment using
  verbal autopsies.} {\em Journal of the American Statistical Association\/}
  111(515):1036--1049.

\bibitem[{McCormick et~al.(2016{\natexlab{b}})McCormick, Li, Calvert, Crampin,
  Kahn and Clark}]{mccormick2016probabilisticSupplement}
---{}---{}---. 2016{\natexlab{b}}. \enquote{Supplemental Material:
  Probabilistic cause-of-death assignment using verbal autopsies.} {\em Journal
  of the American Statistical Association\/} 111(515):1036--1049.

\bibitem[{Miasnikof et~al.(2015)Miasnikof, Giannakeas, Gomes, Aleksandrowicz,
  Shestopaloff, Alam, Tollman, Samarikhalaj and Jha}]{miasnikof2015naive}
Miasnikof, P., V.~Giannakeas, M.~Gomes, L.~Aleksandrowicz, A.~Y. Shestopaloff,
  D.~Alam, S.~Tollman, A.~Samarikhalaj and P.~Jha. 2015. \enquote{Naive Bayes
  classifiers for verbal autopsies: comparison to physician-based
  classification for 21,000 child and adult deaths.} {\em BMC medicine\/}
  13(1):286.

\bibitem[{Mohapatra(2016)}]{mohapatra2016level}
Mohapatra, P.~R. 2016. \enquote{Level of evidence of verbal autopsy.} {\em The
  Lancet Global Health\/} 4(6):e367.

\bibitem[{Murray et~al.(2011)Murray, Lopez, Black, Ahuja, Ali, Baqui, Dandona,
  Dantzer, Das, Dhingra et~al.}]{murray2011population}
Murray, C.~J., A.~D. Lopez, R.~Black, R.~Ahuja, S.~M. Ali, A.~Baqui,
  L.~Dandona, E.~Dantzer, V.~Das, U.~Dhingra et~al. 2011. \enquote{Population
  Health Metrics Research Consortium gold standard verbal autopsy validation
  study: design, implementation, and development of analysis datasets.} {\em
  Population health metrics\/} 9(1):27.

\bibitem[{Murray et~al.(2014)Murray, Lozano, Flaxman, Serina, Phillips,
  Stewart, James, Vahdatpour, Atkinson, Freeman et~al.}]{murray2014using}
Murray, C.~J., R.~Lozano, A.~D. Flaxman, P.~Serina, D.~Phillips, A.~Stewart,
  S.~L. James, A.~Vahdatpour, C.~Atkinson, M.~K. Freeman et~al. 2014.
  \enquote{Using verbal autopsy to measure causes of death: the comparative
  performance of existing methods.} {\em BMC medicine\/} 12(1):5.

\bibitem[{Oti and Kyobutungi({2010})}]{ISI:000208221400021}
Oti, S.~O. and C.~Kyobutungi. {2010}. \enquote{{Verbal autopsy interpretation:
  a comparative analysis of the InterVA model versus physician review in
  determining causes of death in the Nairobi DSS}.} {\em {Population Health
  Metrics}\/} {8}.

\bibitem[{Serina et~al.(2015{\natexlab{a}})Serina, Riley, Stewart, Flaxman,
  Lozano, Mooney, Luning, Hernandez, Black, Ahuja et~al.}]{serina2015shortened}
Serina, P., I.~Riley, A.~Stewart, A.~D. Flaxman, R.~Lozano, M.~D. Mooney,
  R.~Luning, B.~Hernandez, R.~Black, R.~Ahuja et~al. 2015{\natexlab{a}}.
  \enquote{A shortened verbal autopsy instrument for use in routine mortality
  surveillance systems.} {\em BMC medicine\/} 13(1):302.

\bibitem[{Serina et~al.(2015{\natexlab{b}})Serina, Riley, Stewart, James,
  Flaxman, Lozano, Hernandez, Mooney, Luning, Black
  et~al.}]{serina2015improving}
Serina, P., I.~Riley, A.~Stewart, S.~L. James, A.~D. Flaxman, R.~Lozano,
  B.~Hernandez, M.~D. Mooney, R.~Luning, R.~Black et~al. 2015{\natexlab{b}}.
  \enquote{Improving performance of the Tariff Method for assigning causes of
  death to verbal autopsies.} {\em BMC medicine\/} 13(1):291.

\bibitem[{{World Health Organization}(2012)}]{whoStandardTools2012}
{World Health Organization}. 2012. {\em Verbal Autopsy Standards: The 2012 WHO
  verbal autopsy instrument\/}.
  \url{http://www.who.int/healthinfo/statistics/verbalautopsystandards/en/index2.html}.

\bibitem[{{World Health Organization}(2017)}]{whoStandardTools2016}
---{}---{}---. 2017. {\em Verbal Autopsy Standards: The 2016 WHO verbal autopsy
  instrument\/}.
  \url{http://www.who.int/healthinfo/statistics/verbalautopsystandards/en/}.

\end{thebibliography}

\end{document}